# The orbital and superhump periods of the dwarf nova SDSS J093249.57+472523.0

Jeremy Shears, Steve Brady, Shawn Dvorak, Enrique de Miguel, Etienne Morelle, Arto Oksanen and Richard Sabo

**Abstract**

We report unfiltered CCD photometry of the eclipsing dwarf nova SDSS J093249.57+472523.0 obtained during its first confirmed outburst in 2011 March. The outburst amplitude was at least 3.0 magnitudes above mean quiescence and it lasted at least 11 days, although we missed the beginning of the outburst. Superhumps having peak-to-peak amplitude up to 0.3 magnitudes were present during the outburst, thereby establishing it to be a member of the SU UMa family. The mean superhump period was $P_{sh}$ = 0.06814(11) d. Analysis of our measurements of eclipse times of minimum, supplemented with data from other researchers, allowed us to measure the orbital period as $P_{orb}$ = 0.06630354(5) d. The superhump period excess was ε = 0.028(1) which is consistent with of SU UMa systems of similar $P_{orb}$. The FWHM eclipse duration varied between 6 and 13 mins and the eclipse depth was up to 1.6 magnitudes.

**Introduction**

SDSS J093249.57+472523.0 was discovered as US 691 at B=19.4 in the Usher survey of faint blue objects at high galactic latitude which was conducted with the 1.2 m Palomar Schmidt (1). Szkody *et al*. (2), in their search for cataclysmic variables in the Sloan Digital Sky Survey, reported that the object had an unusual spectrum, with very strong doubled helium emission lines and Balmer hydrogen lines. Radial velocity measurements suggested an orbital period, $P_{orb}$, near 0.07 d. Photometry of the system during quiescence by Homer *et al*. (3) revealed eclipses which were 2 magnitude deep, with V ≈ 19.0 out of eclipse. Their analysis of the times of 7 eclipses observed April and May 2004 yielded 3 possible values $P_{orb}$ = 0.0663035(4), 0.0661618(4) or 0.0660206(5) d, with a preference for the middle value. Based on X-ray observations, they suggested it might be an Intermediate Polar.

Observations by the Catalina Real-Time Transient Survey (CRTS) (4) show the system varying between V= 17.5 and 20.5 and with a mean of V=18.8.

The current outburst, the first on record, was detected by JS on 2011 March 18 (5). Follow-up photometry was carried out by a worldwide network of observers using small telescopes. The results of this campaign are presented in this paper

The object is located in UMa at RA 09 32 49.57 Dec +47 25 23.0 (J2000.0).

**Photometry and analysis**

Approximately 76 hours of unfiltered photometry was obtained during the 2011 outburst of SDSS J093249.57+472523.0 using the instrumentation shown in Table 1 and according to the observation log in Table 2. Images were dark-subtracted and flat-fielded prior to being measured using differential aperture photometry relative to star 154 (V= 15.434) on AAVSO chart 5064egs or USNO B1.0 1373-0246596 (V=13.49). Given that the observers used different comparison stars and instrumentation, including CCD cameras with different spectral responses, small systematic differences are likely to exist between observers. However, since the main aim of our research was to look for time dependent phenomena, we do not consider this to be a significant disadvantage. Barycentric corrections were applied to all data.

**Outburst light curve**

The overall light curve of the outburst is shown in Figure 1a. In Figure 2 we plot expanded views of the time series photometry, where each panel shows one day's data drawn to the same scale. This clearly shows recurrent eclipses superimposed on an underlying modulation, which we interpret as superhumps, each of which will be considered in more detail later. The presence of superhumps is diagnostic that SDSS J093249.57+472523.0 is an SU UMa-type dwarf nova, making this the first confirmed superoutburst of the star.

The outburst light curves of most dwarf novae tend to be broadly similar with a plateau phase that lasts for several days, during which there may be a gradual fade, which is followed by a rapid decline towards quiescence. The profile of the outburst of SDSS J093249.57+472523.0 is rather complex by comparison and we have divided in into 4 Sections. The star was at its brightest when it was first detected (mag 15.8). It faded rapidly over the next two days (JD 2455639 to 2455641; indentified as "Section 1" in Figure 1a) at a mean rate of 0.6 mag/d. It then brightened over the next 3 days (JD 2455641 to 2455644; Section 2) at a mean rate of 0.2 mag/d. There was then a slight fade on JD 2455645, and a slight brightening on JD 2455646 (Section 3), before finally fading towards quiescence (Section 4). The outburst was observed for 11 days, but since the star was already fading rapidly on the detection night, it is likely that it had started several days before (we made no observations of the field in the 2 weeks before the detection). Taking the mean CRTS brightness V= 18.8 as mean quiescence, the outburst amplitude was at least 3 magnitudes.

**Measurement of the orbital period**

Times of minimum were measured for the 44 eclipses observed during the outburst using the Kwee and van Woerden method (6) in the *Peranso v2.5* software (7) and are shown in Table 3. In some cases errors were larger due the eclipse being defined by rather few data points. These are supplemented with times from Homer *et*

*al.* (3), Gänsicke (8) and Kato *et al.* (9) (we only used times of eclipses which we did not observe from the latter paper). The orbital period was then calculated from a linear fit to these times of minima as $P_{orb}$ = 0.06630354(5) d. The eclipse time of minimum ephemeris is:

$$BJD_{min} = 2453106.68411(2) + 0.06630354(5) \times E \qquad \text{Equation 1}$$

Our value of $P_{orb}$ is consistent with the shortest of the three possible values proposed by Homer *et al.* (3). as well as the value of 0.066303547(6) d reported by Kato *et al.* (9) in an independent analysis of the 2011 outburst. The O-C (Observed – Calculated) residuals of the eclipse minima relative to the ephemeris in Equation 1 are given in Table 3.

**Measurement of the superhump period**

On the first night of observations, the peak-to-peak amplitude of the superhumps was 0.3 magnitudes. However, analysis of the superhumps during the outburst was complicated by the presence of the eclipses, which often distorted the shape of the superhumps. To study the superhump behaviour, we first extracted the times of each sufficiently well-defined superhump maximum using the Kwee and van Woerden method (6) in *Peranso v2.5* (7). Times of 15 superhump maxima were found and are listed in Table 4. An analysis of the times of maximum for cycles 0 to 105 (JD 2455639 to 2455646) allowed us to obtain the following linear superhump maximum ephemeris:

$$BJD_{max} = 2455639.45618(96) + 0.06814(11) \times E \qquad \text{Equation 2}$$

This gives a mean superhump period during this interval of $P_{sh}$ = 0.06814(11) d. The O–C residuals for the superhump maxima are shown in Figure 1d. The plot suggests that $P_{sh}$ was slightly longer near the beginning than later in the outburst. A separate linear analysis of the superhump times on JD 2455639 to 2455641, at the beginning, gave $P_{sh}$ = 0.06880(20) d). Whereas later in the outburst, between JD 2455644 and 2455646, $P_{sh}$ = 0.06799(17) d. The data are also consistent with a period decrease during the outburst with $dP_{sh}/dt = - 9(3) \times 10^{-5}$.

To confirm our measurements of $P_{sh}$, we carried out a period analysis of the data using the Lomb-Scargle method in *Peranso v2.5* (7). Figure 3a shows the resulting power spectrum of the data from JD 2455639 to 2455646. This has its highest peak at 15.0809(291) cycles/d, which we interpret as the orbital signal, corresponding to $P_{orb}$ = 0.06631(13) d. The error estimates were derived using the Schwarzenberg-Czerny method (10). Such value of $P_{orb}$ is consistent with the one we determined by analysing the eclipses times of minimum. We then pre-whitened the power spectrum to remove the orbital signal which resulted in the power spectrum shown in Figure 3b. In this case the strongest signal was at 14.6686 cycles/d (plus its 1 cycle/d aliases). We interpret this as the superhump signal which corresponds to a superhump period of $P_{sh}$ = 0.06817(9) d which is consistent with the value of $P_{sh}$ derived from the times of superhump analysis.

**Analysis of the eclipses**

We measured the eclipse duration as the full width at half minimum (FWHM; Table 3). Figure 1b shows that the eclipse duration increased from 6 to 13 mins during the first part of the outburst which corresponds to the rapid fade in Section 1 of Figure 1a. Since eclipse duration gives an indication of the size of the accretion disc, this suggests that the accretion disc was expanding during this time. This is in contrast to what is observed in many SU UMa systems, such as SDSS J150240.98+333423.9 (11), where the accretion disc shrinks as the star fades. In the subsequent stages of the outburst, the eclipse duration was about 10 mins. This is above the quiescence eclipse duration of 5 to 6 mins, which we estimate from Figure 6 in Homer *et al.* (3), confirming that the disc was still in an expanded state.

Figure 1c shows that there was also a trend of increasing eclipse depth during the outburst from ~0.9 mag near the beginning to ~1.6 mag towards the end (data are given in Table 3), but still above the 2.0 magnitude quiescence eclipses in Homer *et al.* (3). However, a cursory examination of the light curves presented in Figure 2 shows that the eclipse depth is also affected by the location of the superhump: in general eclipses are shallower when hump maximum coincides with eclipse. This explains the cluster of relatively shallow eclipses observed around JD 2455641.Such variation in eclipse depth is commonly observed in SU UMa systems including DV UMa, IY UMa and SDSS J122740.82+513925.0 and SDSS J150240.98+333423.9 (12) (13) (14) (11), where there is a relationship between eclipse depth and the precession period of the accretion disc, $P_{prec}$, sometimes referred to as the beat period of the superhump and orbital periods.

According to the relation $1/P_{prec} = 1 / P_{orb} - 1/ P_{sh}$, the precession period should be about 2.46(15) d, based on our measured values of $P_{sh}$ and $P_{orb}$. We performed a Lomb-Scargle analysis on the combined data in the interval 0.1 to 1 cycles/d (1 to 10 d). The resulting power spectrum in Figure 4 shows, in order of strength, the following 3 signals:

| | | |
|---|---|---|
| Peak 1 | 0.60(3) c/d | 1.68(8) d |
| Peak 2 | 0.80(5) c/d | 1.24(8) d |
| Peak 3 | 0.42(4) c/d | 2.40(24) d |

It is tempting to associate Peak 3 with $P_{prec}$ and, possibly, Peak 2 with $P_{prec}$ /2. However, none of the signals were particularly strong, hence caution should be exercised in their interpretation.

Given that the accretion disc is elliptical during a superoutburst, the angle it subtends will vary as it precesses as viewed from the Earth. This itself could cause the duration of the eclipse of the accretion disc to vary. We have considered the possibility, therefore, that the increase in eclipse duration observed in Section 1 of the outburst light curve, and discussed above, is due to this effect, rather than an inherent change in the dimensions of the disc. However we note the increase was observed over an interval of 2 days, which is a large fraction of the proposed precession period, hence this must remain speculation for the present. Moreover the ratio of the duration of the eclipse varied by a factor of 2, which would imply a highly elliptical disc, which seems unlikely.

**Estimation of the secondary to primary mass ratio**

Taking our measured orbital period, $P_{orb}$ = 0.06630354(5) d and our mean superhump period of $P_{sh}$ = 0.06814(11) d, we calculate the superhump period excess ε = 0.028(1). Such value is consistent with other SU UMa systems of similar orbital period (15)

Patterson *et al.* (16) established an empirical relationship between ε and q, the secondary to primary mass ratio: ε = 0.18*q + 0.29*$q^2$. This assumes a white dwarf of ~0.75 solar masses which is typical of SU UMa systems. Our value of ε = 0.028 allows us to estimate q = 0.16.

**Discussion**

As noted previously, SDSS J093249.57+472523.0 has a rather unusual outburst light curve which is not typical for SU UMa systems. Moreover, quiescence photometry by Homer *et al.* (3) a higher structure than typical SU UMa systems. In this regards it resembles the Intermediate Polar HT Cam, which shows a fair amount of variability during quiescence. We also note that the spectrum shown by Szkody *et al.* is rather unusual amongst SU UMa systems, with strong HeI/II emission lines. One could speculate that if indeed SDSS J093249.57+472523.0 is an Intermediate Polar, or some other weakly magnetic system, this could affect the shape of the (super)outburst light curve, causing the unusual dip that we observed.

This unusual system warrants further study. Homer *et al.* (3) reported a second periodicity in their optical light curves at quiescence, P= 0.0688 (+0.0012-0.0006) d, which is significantly longer than $P_{orb}$. Although this detection was based on a relatively small amount of data, it may indicate the presence of permanent superhumps and further observations during quiescence would confirm or refute this idea.

**Conclusions**

Our observations of the first recorded outburst of SDSS J093249.57+472523.0 showed that the amplitude was at least 3.0 magnitudes above mean quiescence and it lasted at least 11 days. Superhumps having peak-to-peak amplitude up to 0.3 magnitudes were present, thereby establishing it to be a member of the SU UMa family. The mean superhump period was $P_{sh}$ = 0.06814(11) d. Analysis of the eclipse times of minimum, supplemented with data from other researchers, allowed us to measure the orbital period as $P_{orb}$ = 0.06630354(5) d. The superhump period excess was ε = 0.028(1), from which we estimated the secondary to primary mass ratio, q = 0.16. The FWHM eclipse duration varied between 6 and 13 mins and the eclipse depth was up to 1.6 magnitudes.

We encourage further monitoring of SDSS J093249.57+472523.0 with the aim of identifying further outbursts. Photometry during a future outburst may shed more light on the unusual light curve and other behaviour of the system which we observed in the 2011 outburst.

## Acknowledgements

The authors gratefully acknowledge the use of data from the Catalina Real-Time Transient Survey. We used SIMBAD, operated through the Centre de Données Astronomiques (Strasbourg, France), and the NASA/Smithsonian Astrophysics Data System. We are indebted to our referees, Professor Boris Gänsicke (Warwick University, UK) and Professor Coel Hellier (Keele University, UK) for their helpful comments that have improved the paper.

**Addresses**

JS: "Pemberton", School Lane, Bunbury, Tarporley, Cheshire, CW6 9NR, UK [bunburyobservatory@hotmail.com]

SB: 5 Melba Drive, Hudson, NH 03051, USA [sbrady10@verizon.net]

SD: Rolling Hills Observatory, Clermont, FL, USA [sdvorak@rollinghillsobs.org]

EdM: Departamento de Fisica Aplicada, Facultad de Ciencias Experimentales, Universidad de Huelva, 21071 Huelva, Spain; Center for Backyard Astrophysics, Observatorio del CIECEM, Parque Dunar, Matalascañas, 21760 Almonte, Huelva, Spain [demiguel@uhu.es]

EM: Lauwin-Planque Observatory, F-59553 Lauwin-Planque, France [etmor@free.fr]

AO: Hankasalmi observatory, Verkkoniementie 30, FI-40950 Muurame, Finland [arto.oksanen@jklsirius.fi]

RS: 2336 Trailcrest Dr., Bozeman, MT 59718, USA [richard@theglobal.net]

| Observer | Telescope | CCD |
|---|---|---|
| Brady | 0.4 m reflector | SBIG ST-8XME |
| Dvorak | 0.25 m SCT | SBIG ST-9XE |
| de Miguel | 0.25 m reflector | QSI-516ws |
| Morelle | 0.4 m SCT | SBIG ST-9 |
| Oksanen | 0.4 m reflector | SBIG STL-1001E |
| Sabo | 0.43 m reflector | SBIG STL-1001 |
| Shears | 0.28 m SCT | Starlight Xpress SXVF-H9 |

**Table 1: Instrumentation**

| Start time in 2011 (UT) | Start time (JD) | End time (JD) | Duration (h) | Observer |
|---|---|---|---|---|
| March 18 | 2455639.400 | 2455639.529 | 3.1 | Shears |
| March 19 | 2455639.571 | 2455639.684 | 2.7 | de Miguel |
| March 19 | 2455640.333 | 2455640.673 | 8.1 | de Miguel |
| March 20 | 2455640.549 | 2455640.855 | 7.3 | Brady |
| March 20 | 2455640.638 | 2455640.850 | 5.1 | Dvorak |
| March 20 | 2455640.734 | 2455640.990 | 6.1 | Sabo |
| March 20 | 2455641.328 | 2455641.660 | 8.0 | de Miguel |
| March 21 | 2455642.329 | 2455642.414 | 2.0 | de Miguel |
| March 22 | 2455643.367 | 2455643.498 | 3.1 | Morelle |
| March 23 | 2455644.324 | 2455643.627 | 7.3 | Morelle |
| March 24 | 2455645.309 | 2455645.620 | 7.5 | Morelle |
| March 25 | 2455646.282 | 2455646.629 | 8.3 | Morelle |
| March 26 | 2455647.426 | 2455647.450 | 0.6 | Oksanen |
| March 27 | 2455648.351 | 2455648.358 | 0.1 | Shears |
| March 29 | 2455650.311 | 2455650.592 | 6.7 | Oksanen |
| April 3 | 2455655.362 | 2455655.370 | 0.1 | Shears |

**Table 2: Log of time-series observations**

| Orbit number | Eclipse minimum (BJD) | Uncertainty (d) | O-C (d) | Eclipse depth (mag) | Eclipse duration (min) | Source |
|---|---|---|---|---|---|---|
| 0 | 2453106.6845 | 0.0005 | 0.0004 | | | Ref. (3) |
| 1 | 2453106.7504 | 0.0000 | 0.0000 | | | Ref. (3) |
| 2 | 2453106.8169 | 0.0000 | 0.0002 | | | Ref. (3) |
| 467 | 2453137.6482 | 0.0002 | 0.0003 | | | Ref. (3) |
| 468 | 2453137.7142 | 0.0001 | 0.0000 | | | Ref. (3) |
| 469 | 2453137.7801 | 0.0002 | -0.0004 | | | Ref. (3) |
| 470 | 2453137.8470 | 0.0003 | 0.0002 | | | Ref. (3) |
| 4102 | 2453378.6611 | 0.0000 | -0.0001 | | | Ref. (8) |
| 4103 | 2453378.7274 | 0.0000 | -0.0002 | | | Ref. (8) |
| 4104 | 2453378.7935 | 0.0000 | -0.0003 | | | Ref. (8) |
| 38199 | 2455639.4148 | 0.0006 | -0.0001 | 1.06 | 5.5 | |
| 38200 | 2455639.4800 | 0.0030 | -0.0002 | 0.90 | 7.2 | |
| 38202 | 2455639.6124 | 0.0006 | -0.0003 | 1.18 | ND | |
| 38203 | 2455639.6796 | 0.0006 | 0.0017 | 0.90 | 6.8 | |
| 38209 | 2455640.0745 | 0.0001 | 0.0006 | | | Ref. (9) |
| 38210 | 2455640.1419 | 0.0001 | 0.0004 | | | Ref. (9) |
| 38211 | 2455640.2085 | 0.0001 | 0.0013 | | | Ref. (9) |
| 38213 | 2455640.3424 | 0.0009 | -0.0016 | 0.53 | 8.6 | |
| 38214 | 2455640.406 | 0.0003 | -0.0005 | 0.71 | ND | |
| 38215 | 2455640.4759 | 0.0003 | -0.0002 | 0.59 | ND | |
| 38216 | 2455640.54 | 0.0009 | 0.0011 | 0.73 | 7.2 | |
| 38217 | 2455640.6045 | 0.0009 | -0.0017 | 0.66 | 11.5 | |
| 38217 | 2455640.604 | 0.0030 | 0.0020 | 0.59 | ND | |
| 38217 | 2455640.605 | 0.0018 | -0.0006 | 0.59 | ND | |
| 38218 | 2455640.6727 | 0.0018 | -0.0020 | 0.80 | ND | |
| 38218 | 2455640.6717 | 0.0009 | -0.0022 | 0.72 | 10.1 | |
| 38218 | 2455640.6717 | 0.0009 | -0.0014 | 0.74 | ND | |
| 38218 | 2455640.6720 | 0.0009 | -0.0001 | 0.75 | ND | |
| 38219 | 2455640.7374 | 0.0009 | -0.0011 | 0.76 | ND | |
| 38219 | 2455640.7369 | 0.0015 | -0.0011 | 0.76 | ND | |
| 38219 | 2455640.7370 | 0.0012 | -0.0008 | 0.75 | ND | |
| 38220 | 2455640.8056 | 0.0012 | -0.0017 | 0.74 | ND | |
| 38220 | 2455640.8058 | 0.0015 | -0.0022 | 0.75 | ND | |
| 38227 | 2455641.2694 | 0.0001 | -0.0021 | | | Ref. (9) |
| 38228 | 2455641.3378 | 0.0003 | 0.0002 | 1.02 | 11.7 | |
| 38229 | 2455641.4049 | 0.0006 | 0.0004 | 1.19 | 13.4 | |
| 38232 | 2455641.6018 | 0.0003 | 0.0001 | 1.13 | 12.5 | |
| 38244 | 2455642.3977 | 0.0006 | 0.0019 | 1.19 | ND | |
| 38259 | 2455643.3913 | 0.0009 | 0.0027 | 1.28 | 9.4 | |
| 38260 | 2455643.4573 | 0.0012 | 0.0007 | 1.14 | 7.8 | |
| 38274 | 2455644.3879 | 0.0018 | 0.0010 | ND | ND | |
| 38275 | 2455644.4532 | 0.0021 | 0.0000 | ND | ND | |
| 38276 | 2455644.5201 | 0.0006 | -0.0003 | 1.29 | 8.9 | |

| 38277 | 2455644.5866 | 0.0012 | 0.0021 | 1.41 | 9.9 | |
| --- | --- | --- | --- | --- | --- | --- |
| 38289 | 2455645.3784 | 0.0033 | 0.0011 | ND | ND | |
| 38290 | 2455645.4467 | 0.0006 | 0.0017 | 0.78 | 11.7 | |
| 38291 | 2455645.5132 | 0.0006 | 0.0019 | 0.87 | 8.9 | |
| 38292 | 2455645.5795 | 0.0006 | -0.0020 | 1.04 | 8.6 | |
| 38303 | 2455646.3096 | 0.0003 | 0.0000 | 1.23 | 11.7 | |
| 38304 | 2455646.3774 | 0.0009 | 0.0002 | 1.06 | 9.4 | |
| 38305 | 2455646.4422 | 0.0006 | 0.0002 | 1.20 | 10.4 | |
| 38306 | 2455646.5091 | 0.0003 | 0.0010 | 1.25 | 8.9 | |
| 38307 | 2455646.5756 | 0.0018 | 0.0025 | 1.10 | 9.8 | |
| 38314 | 2455647.0375 | 0.0001 | 0.0010 | | | Ref. (9) |
| 38315 | 2455647.1035 | 0.0001 | 0.0016 | | | Ref. (9) |
| 38319 | 2455647.3689 | 0.0000 | 0.0018 | | | Ref. (9) |
| 38320 | 2455647.4365 | 0.0009 | -0.0004 | 1.34 | 8.4 | |
| 38328 | 2455647.9649 | 0.0001 | -0.0007 | | | Ref. (9) |
| 38329 | 2455648.0318 | 0.0001 | -0.0006 | | | Ref. (9) |
| 38330 | 2455648.0987 | 0.0000 | 0.0007 | | | Ref. (9) |
| 38364 | 2455650.3529 | 0.0009 | -0.0013 | 1.56 | 9.1 | |
| 38365 | 2455650.4198 | 0.0006 | -0.0007 | 1.41 | 7.5 | |
| 38366 | 2455650.4864 | 0.0009 | -0.0001 | 1.60 | 8.8 | |
| 38367 | 2455650.5523 | 0.0006 | -0.0002 | 1.43 | 8.5 | |

**Table 3: Eclipse minimum times, depth and duration**
Data from the present study, unless referenced otherwise. ND = not determined

| Superhump cycle number | Superhump maximum (BJD) | Uncertainty (d) | O-C (d) |
| --- | --- | --- | --- |
| 0 | 2455639.4527 | 0.0015 | -0.0035 |
| 1 | 2455639.5208 | 0.0045 | -0.0035 |
| 2 | 2455639.5831 | 0.0010 | -0.0094 |
| 3 | 2455639.6611 | 0.0030 | 0.0005 |
| 14 | 2455640.4155 | 0.0015 | 0.0054 |
| 15 | 2455640.4852 | 0.0030 | 0.0069 |
| 16 | 2455640.5496 | 0.0010 | 0.0032 |
| 72 | 2455644.3635 | 0.0060 | 0.0012 |
| 73 | 2455644.4342 | 0.0055 | 0.0038 |
| 75 | 2455644.5699 | 0.0075 | 0.0032 |
| 101 | 2455646.3376 | 0.0015 | -0.0007 |
| 102 | 2455646.4048 | 0.0020 | -0.0017 |
| 103 | 2455646.4728 | 0.0015 | -0.0018 |
| 104 | 2455646.5415 | 0.0010 | -0.0012 |
| 105 | 2455646.6083 | 0.0025 | -0.0026 |

**Table 4: Superhump maximum times**

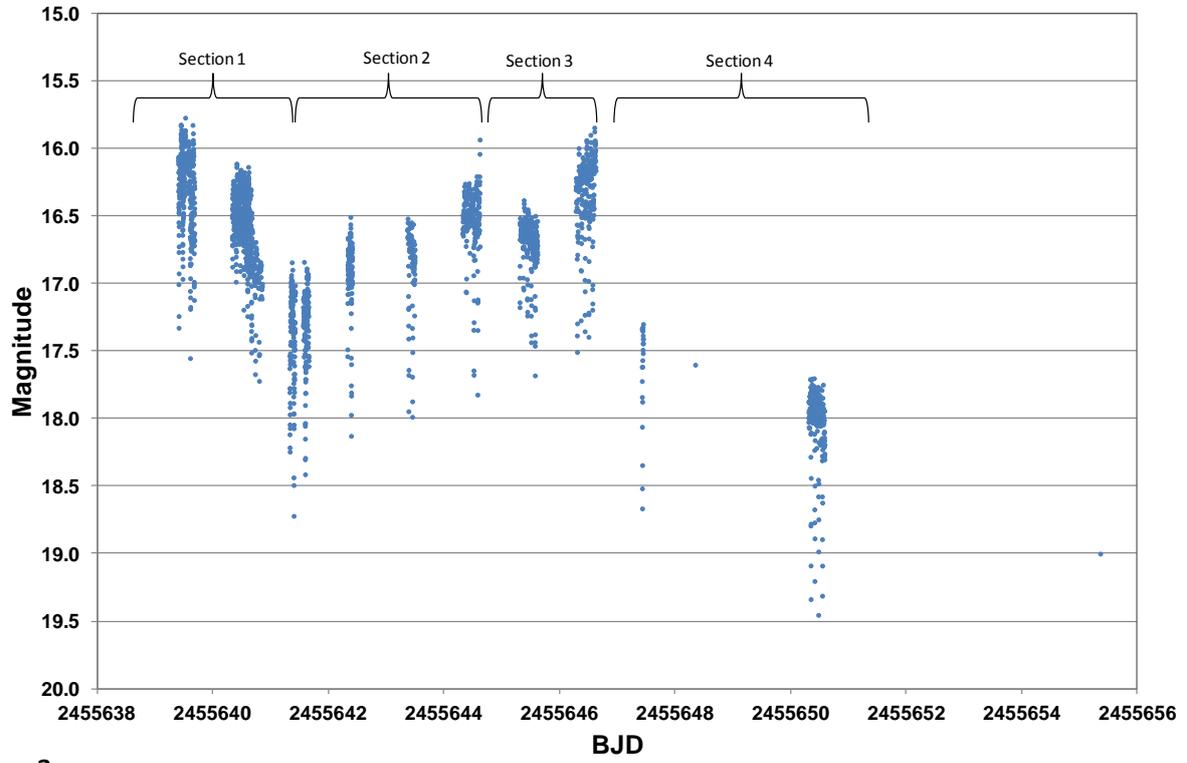
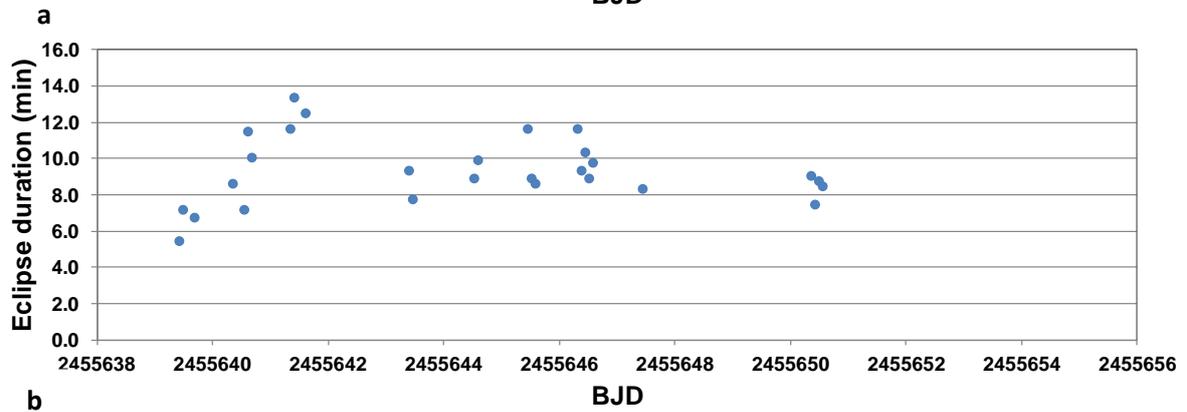
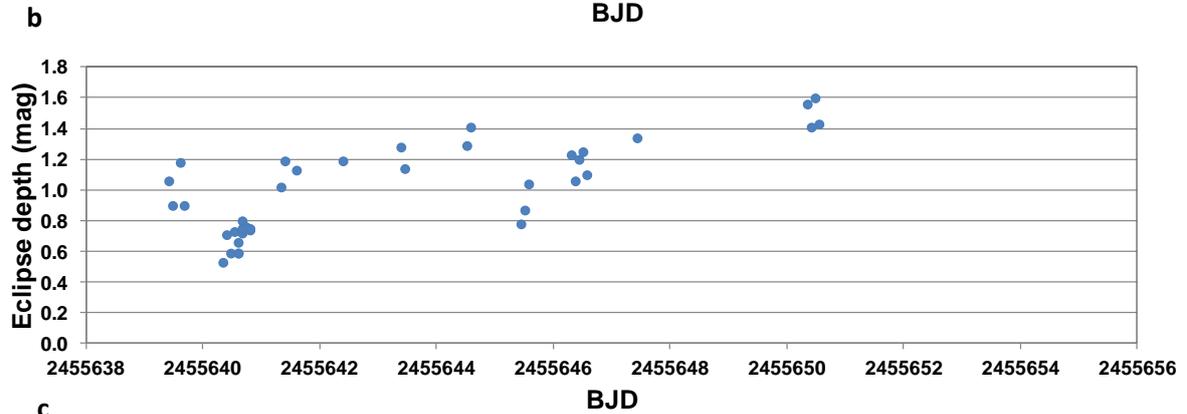
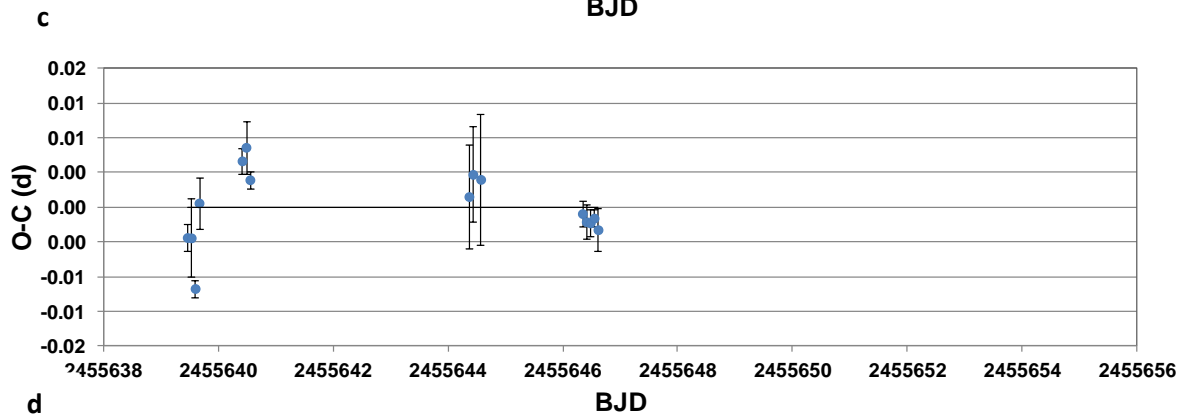

Figure 1 (a) outburst light curve, (b) eclipse duration, (c) eclipse depth (d) superhump O-C diagram

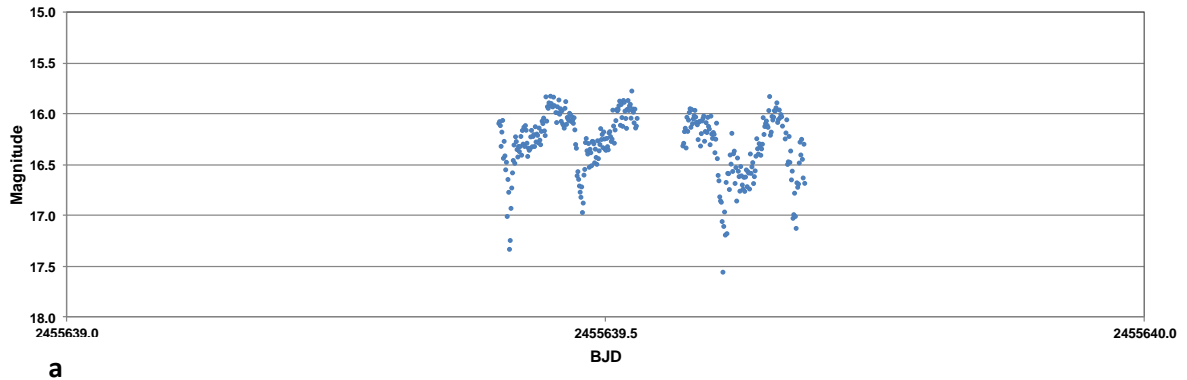

a

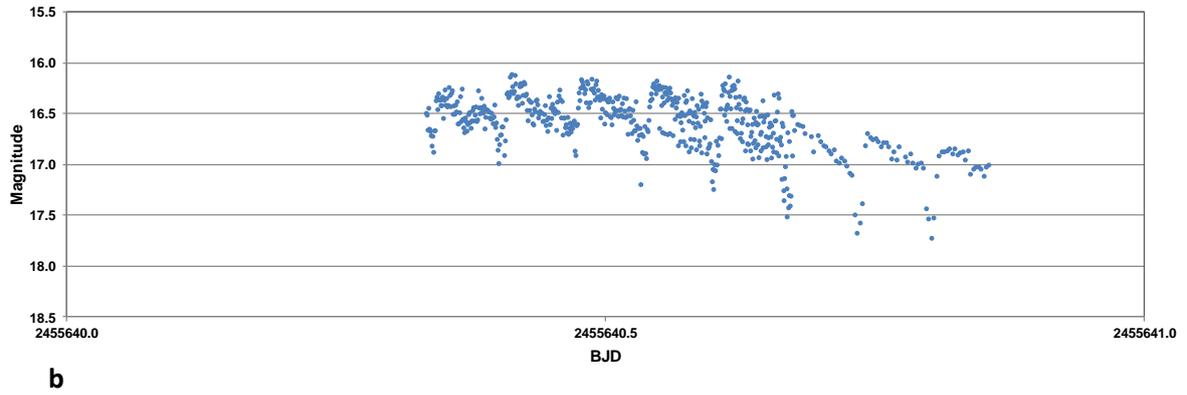

b

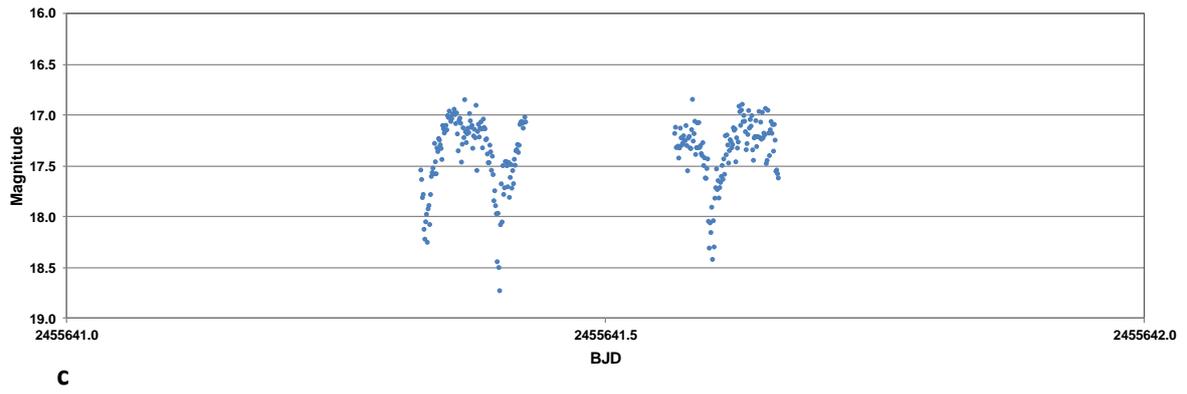

c

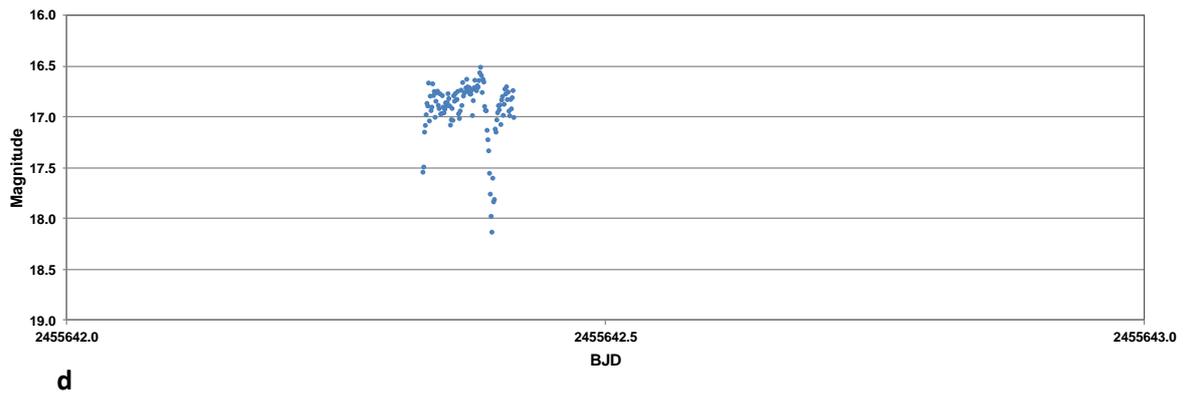

d

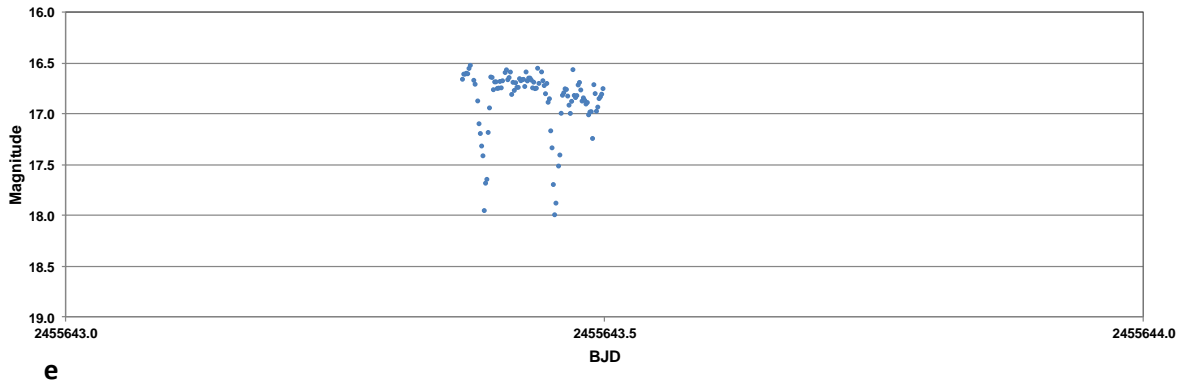

e

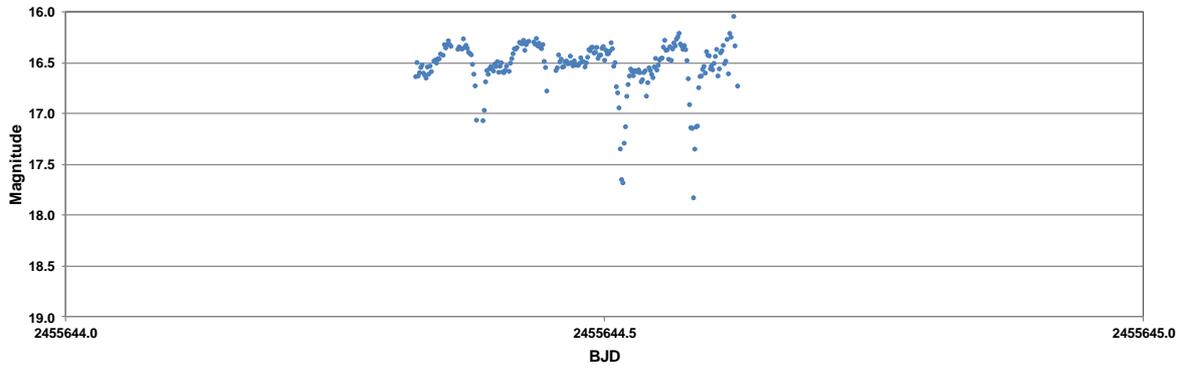

f

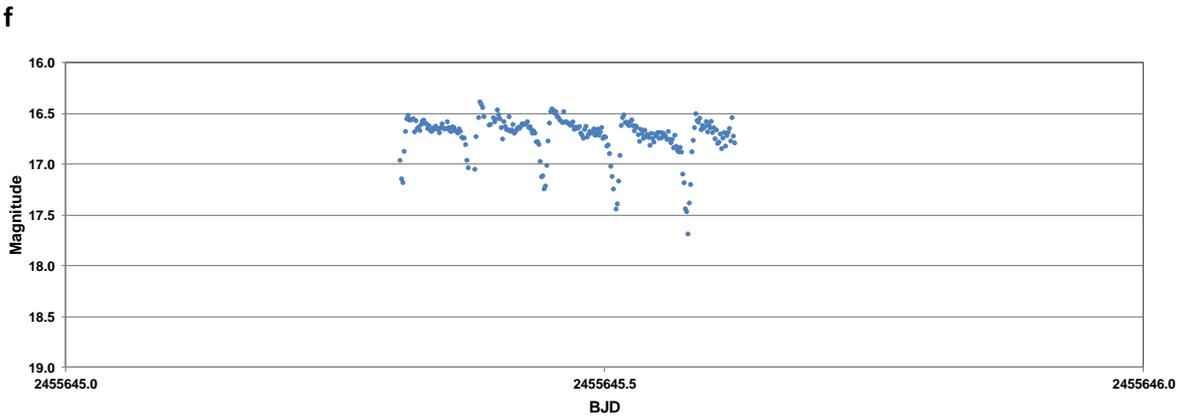

g

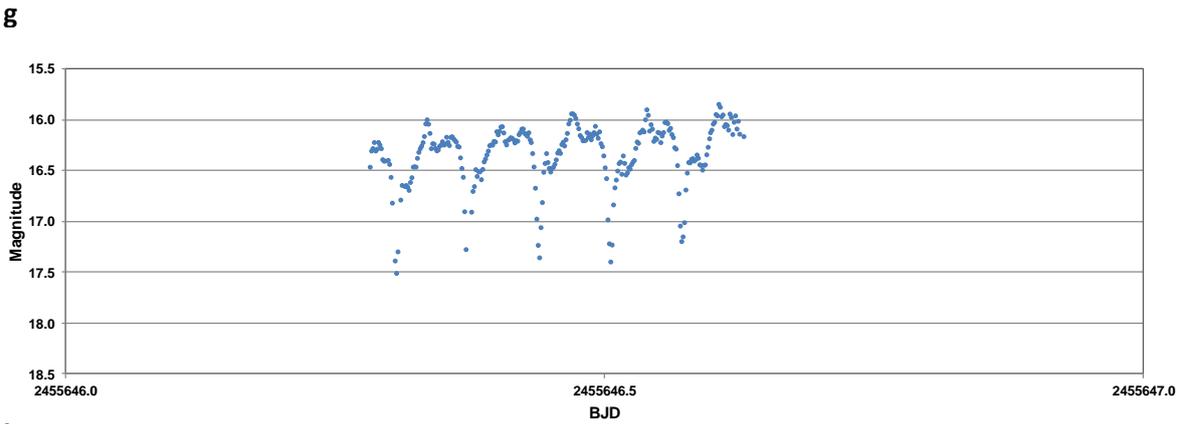

h

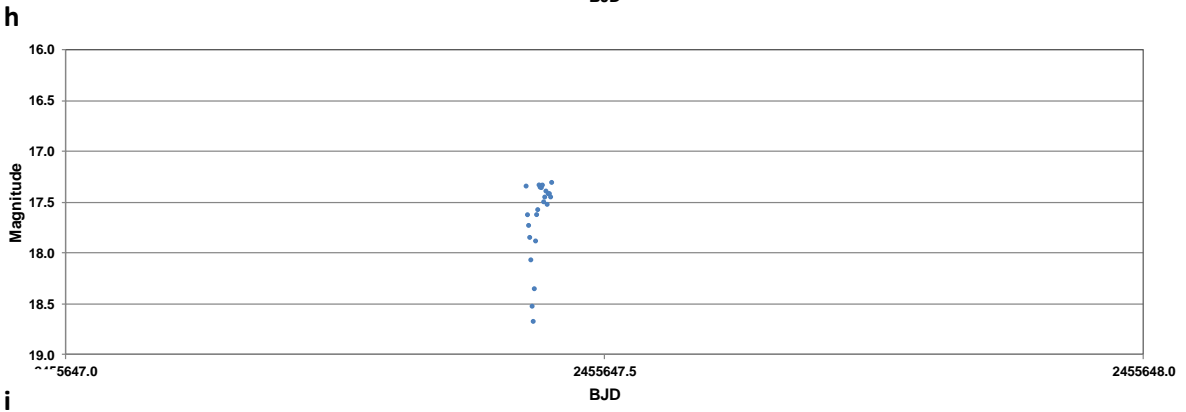

i

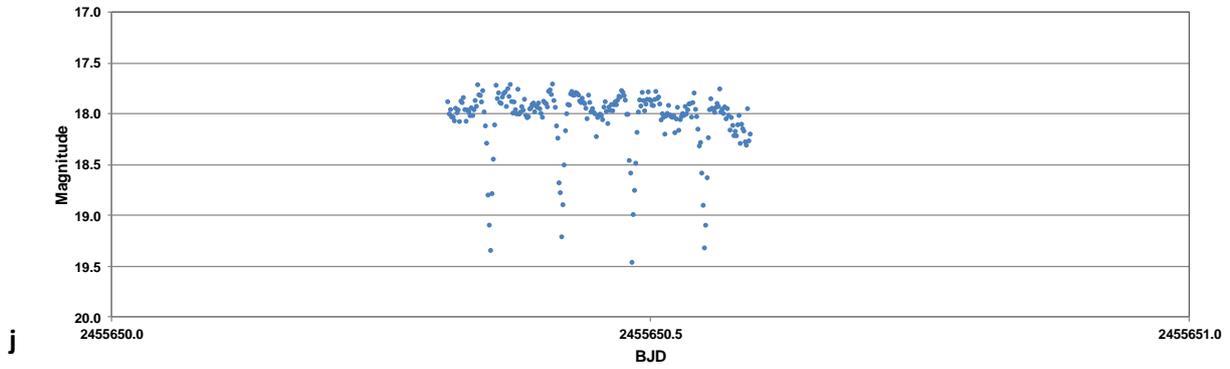

j

Figure 2: Expanded views of the time series photometry during the outburst

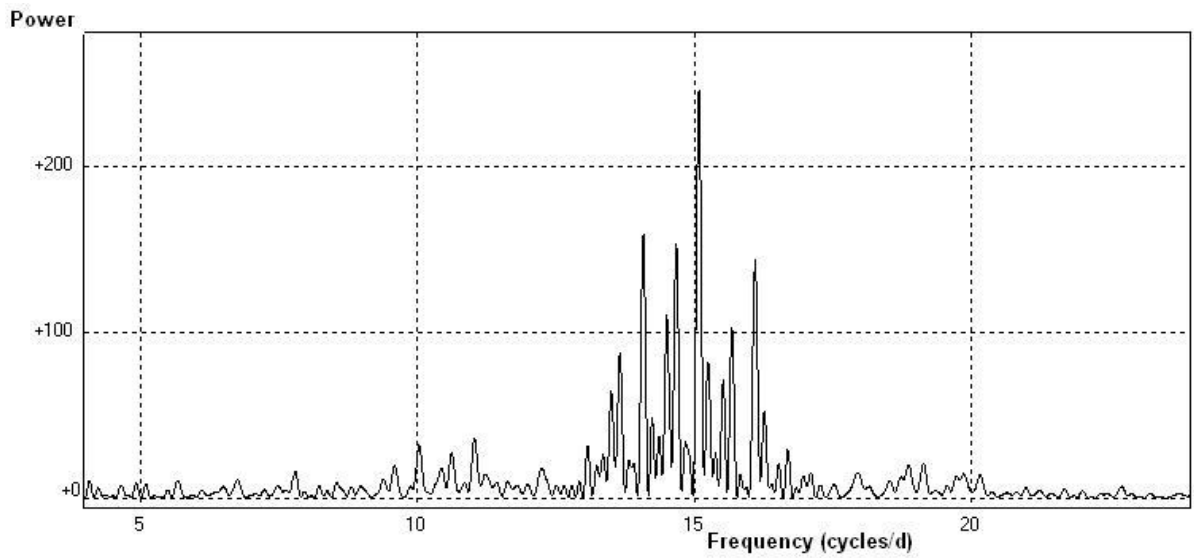

a

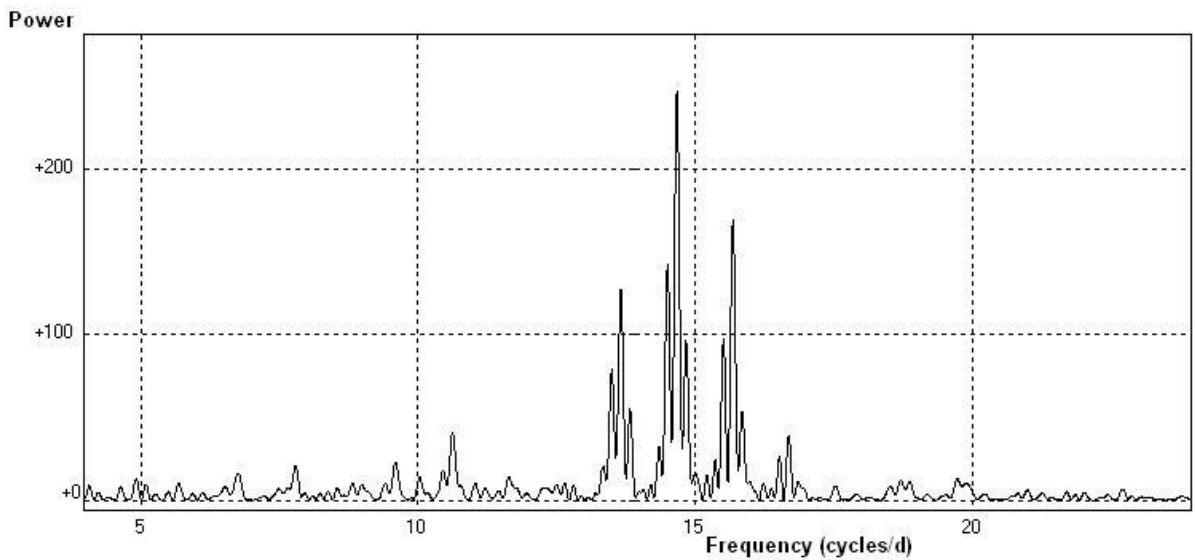

b

Figure 3: (a) Lomb-Scargle power spectrum of the data from JD 2455639 to 2455646 in the interval 4 to 24 cycles/d, (b) Lomb-Scargle spectrum after pre-whitening with the orbital signal (15.0809 cycles/d)

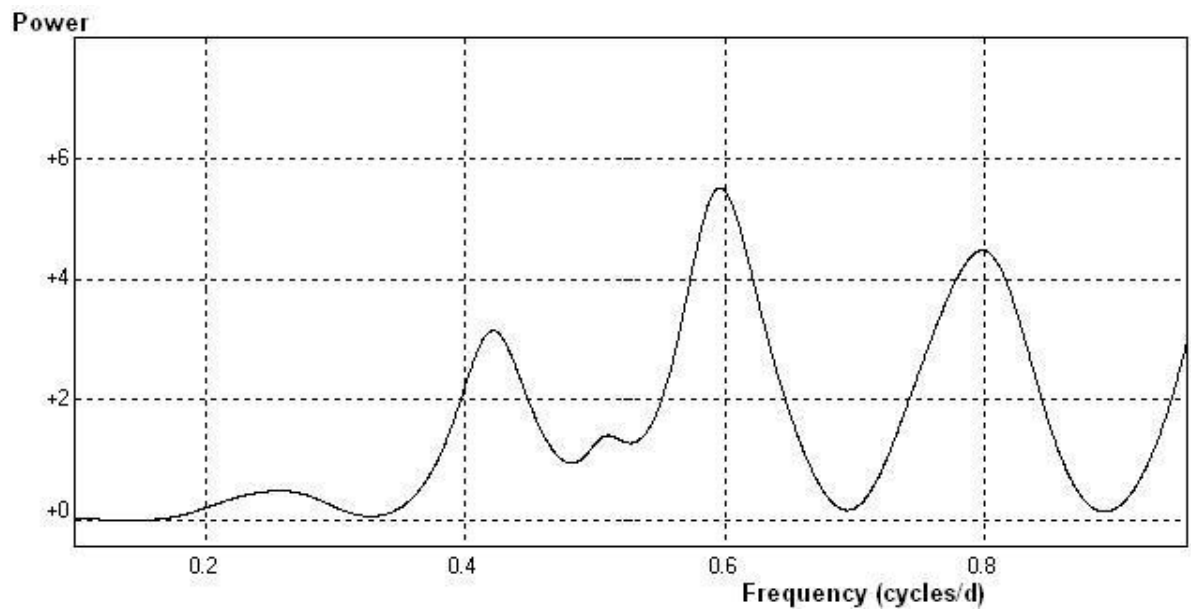

**Figure 4: Lomb-Scargle power spectrum of the data from JD 2455639 to 2455646 in the interval 0.1 to 1.0 cycles/d**